\documentclass[]{IEEEtran}

\usepackage{amsmath}
\usepackage{booktabs}

\input epsf
\usepackage{graphicx}
\usepackage{wrapfig}
\usepackage{dcolumn}
\usepackage{url}
\usepackage{upgreek}
\usepackage{soul,xcolor}
\usepackage{siunitx}
\usepackage{upgreek,textcomp}
\usepackage{color}
\usepackage[colorinlistoftodos]{todonotes}

\begin{document}

\IEEEoverridecommandlockouts

\title{\LARGE The Performance of the KIBB-g1 Tabletop Kibble Balance at NIST}

\setstcolor{red}
\newcommand{\sss}{\textcolor{blue}}

\author{\IEEEauthorblockN{Leon Chao,
Frank Seifert, Darine Haddad, Jon Pratt, David Newell, and
Stephan Schlamminger} \\
\IEEEauthorblockA{National Institute of Standards and Technology, 100 Bureau Dr., Gaithersburg MD 20899\\leon.chao@nist.gov} \\

%ORCIDs
% Leon: 0000-0001-7589-4019
% Frank:
% Darine:
% Jon: 
% Dave: 0000-0002-2612-1172
% Stephan: 0000-0002-9270-4018

}

\maketitle

\begin{abstract}
%On November 16, 2018, the 26th General Conference on Weights and Measures (CGPM) voted unanimously to revise the International System of Units (SI) from a system  built on seven base units to one built on seven defining constants. The revised SI officially became effective on May 20, 2019, World Metrology Day. The SI unit of mass, the kilogram, is now realized via a fixed value of the Planck constant $h$. Over the past few decades, national metrology institutes around the world have developed Kibble Balances (KB) as a primary SI realization of mass, with the majority aimed at the 1-kg level with uncertainties on the order of a few parts in $10^8$. However, upon fixing the Planck constant, mass can be directly realized at any level, deeming the kilogram only a historically unique benchmark. 

A tabletop-sized Kibble balance (KIBB-g1) designed to directly realize mass at the gram-level range with uncertainties on the order of parts in 10$^6$ has been developed at the National Institute of Standards and Technology (NIST). The masses of a nominally 5\,g and 1\,g weight were determined with 1-$\sigma$ standard uncertainties of 9.0\,$\upmu$g  and 6.7\,$\upmu$g, respectively. The corresponding relative uncertainties are $1.8\times 10^{-6}$ and $6.3\times 10^{-6}$. The construction of the instrument, capabilities, and full uncertainty budgets are presented in this manuscript.
\end{abstract}


\IEEEpeerreviewmaketitle
\pagenumbering{gobble}

\section{Introduction}

The maximum permitted uncertainties for International Organization of Legal Metrology (OIML) class $\mathrm{E_2}$ weights, typically used for calibrating weighing instruments, ranging from 1\,g to 10\,g are from 10\,$\mathrm{\upmu g}$ to 20\,$\mathrm{\upmu g}$, respectively~\cite{OIML}. It is required for these weights to undergo scheduled calibrations, especially when mishandled, by accredited metrology institutes which is a financially and time consuming procedure. With the revised International System (SI) of units, mass can be directly realized at any scale point (i.e. milligram, gram, kilogram, etc.) \cite{Haddad}. Instrument manufacturers and metrology institutes have shown interest in directly measuring small masses with a tabletop Kibble balance (KB) capable of realizing the unit of mass with the same level of uncertainties associated with a set of calibration weights. Operating at this level of relative uncertainty also relaxes the demand for direct connection to quantum electrical standards, gravimeters, and high vacuum environments required for more accurate KBs. Following the initial performance evaluation, \cite{Chao}, we describe design enhancements of KIBB-g1, or (KIB)ble (B)alance at the (g)ram level, version (1), and its capabilities along with full uncertainty budgets for a 1\,g and 5\,g mass determination.

\section{Theory of a Kibble Balance}

A conventional beam balance makes relative measurements, comparing the weight of an object to that of a calibrated mass. A Kibble balance, however, makes absolute measurements, comparing the weight of an object to a frequently calibrated electromagnetic force determined by electrical quantities. The experiment involves two modes of operation, velocity mode and force mode. Velocity mode is based on the principle of Faraday's law of induction. A coil of wire length $L$ is moved at a vertical velocity $v$ through a magnetic field of flux density $B$ so that a voltage $V$ is induced. The induced voltage is related to the velocity through the flux integral $BL$: 
\begin{equation}
V=BLv
\end{equation}

Force mode is based on the Lorentz force. The gravitational force on a mass $m$ is counteracted by an upward electromagnetic force $F$ generated by the same coil, now energized with a current $I$ in the same magnetic field:
\begin{equation}
F= BLI = mg
\end{equation}
where $g$ is the local gravitational acceleration.

An expression that virtually equates electrical and mechanical power leading to a solution for mass is obtained by combining equations (1) and (2):
\begin{equation}
VI=mgv\; \Longrightarrow \;m = \frac{VI}{gv} \label{eq:watt}
\end{equation}
Since KIBB-g1 strives for relative uncertainties on the order of a few parts in $10^6$, the Planck constant only makes a subtle appearance as the means for absolutely calibrating the voltmeter and resistance standard used for the electrical measurements. For a detailed description of KB theory, see \cite{Robinson,Haddad2}.

\section{Design Overview}
\subsection{Mechanical}

The KIBB-g1 Kibble balance design aims to provide industrial laboratories the option to directly measure mass artifacts at the gram level on site without access to any calibrated masses. Listed in order of priority, five design goals were established:

\begin{samepage}
\begin{enumerate}
\item  Nominal values: between  1\,g to 10\,g
\item Relative uncertainties: $\approx 10^{-6}$
\item  Form factor: `tabletop' sized instrument
\item  Convenience: operates in air (no vacuum required)
\item  Cost:  $<$ 50,000 USD
\end{enumerate}
\end{samepage}

KIBB-g1 measures 57\,cm tall and fits on a 30\,cm diameter circular optical breadboard. It is designed such that the `main mass side' (MMS) contains mostly all the components relevant to velocity and force mode while the `counter mass side' (CMS) serves as a driving motor for velocity mode \cite{Chao}. 

\begin{figure}
\center
\includegraphics[width=3.6in]{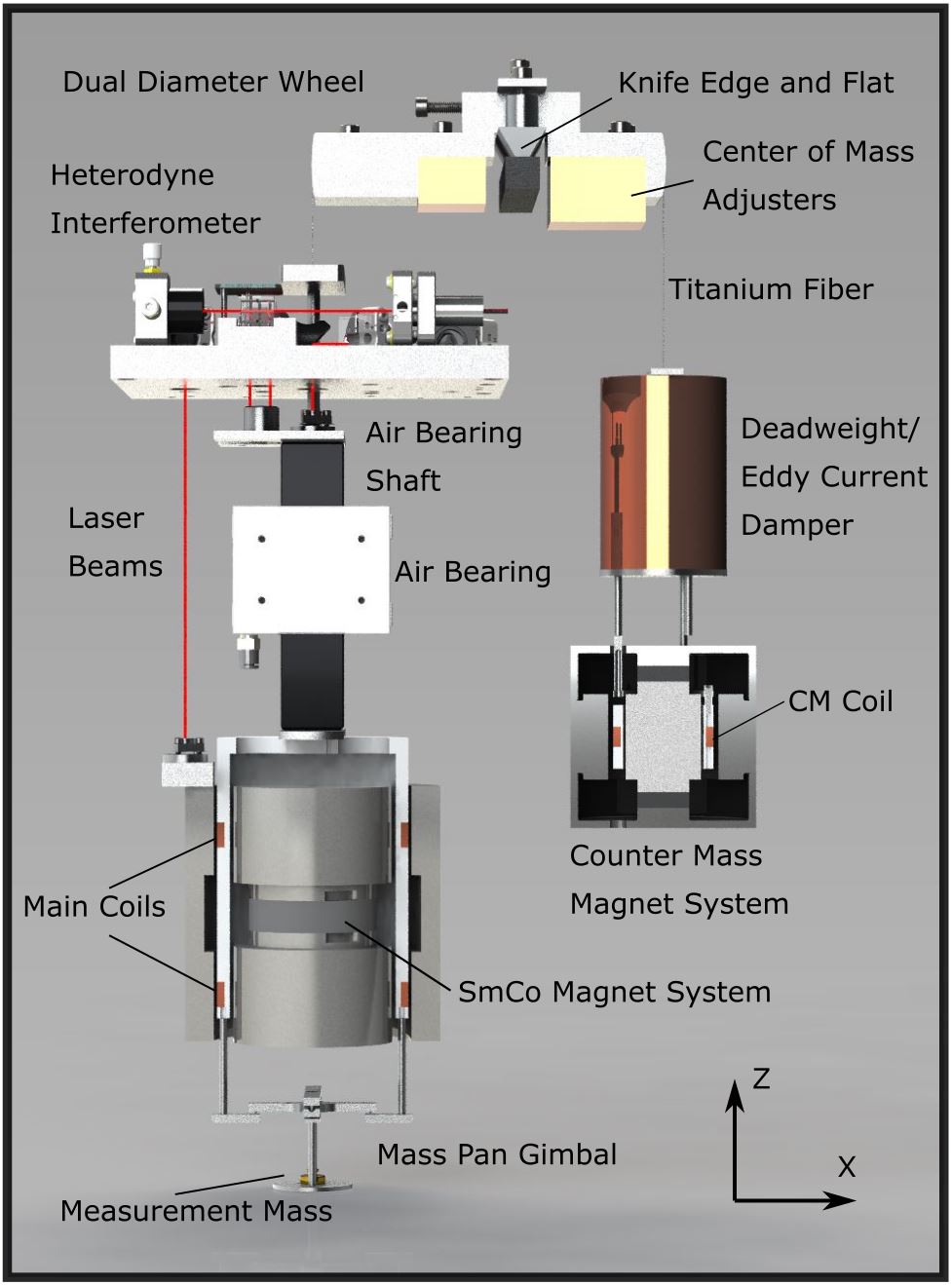}
\caption{CAD rendering of the KIBB-g1 Kibble balance. Structural components are hidden for clarity. Cross-sectional views of both magnets/coils are shown. The main mass side is to the left and the counter mass side to the right  of the knife edge.}
\label{fig:CAD}
\end{figure}

\begin{figure}
\center
\includegraphics[width=3.4in]{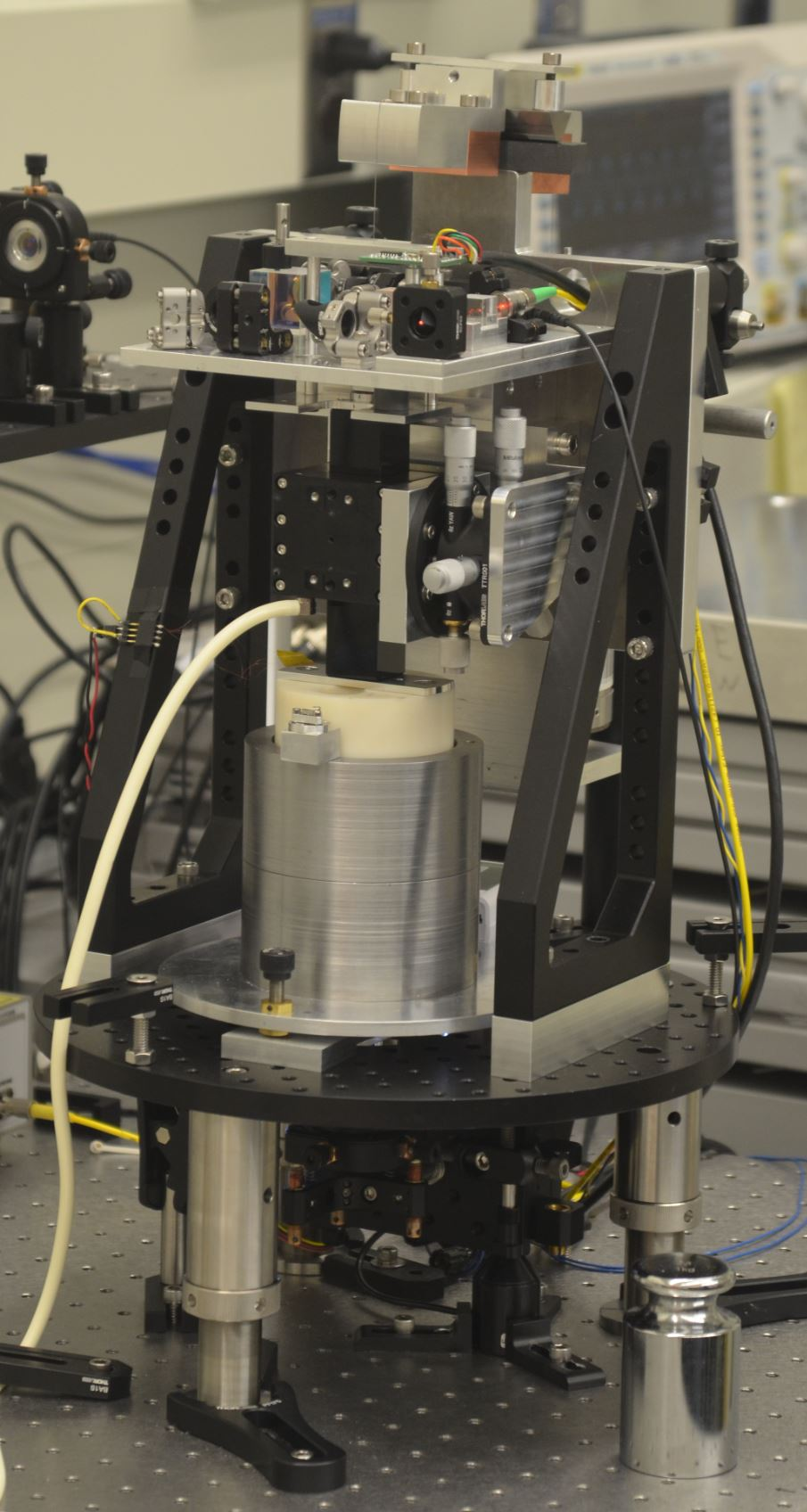}
\caption{Photograph of the KIBB-g1 tabletop Kibble balance, perspective view of the main mass side. White hose supplies the air for the air bearing. The total height of the instrument is 57\,cm tall.}
\label{fig:Photo}
\end{figure}

Starting from the top of the balance as shown in figs. \ref{fig:CAD}--\ref{fig:Photo}, a dual-diameter truncated wheel pivots about a precision ground, tungsten carbide knife edge resting on a diamond-like carbon coated tungsten carbide surface. The truncated wheel looks like a beam, but effectively behaves as a pulley. The  prescribed motion of the hanging coils along the $Z$-axis is  constrained by the rotation of the wheel about the $Y$-axis. The MMS beam arc has a smooth, curved surface with a radius 1.4 times that of the CMS arc. Both arcs are centered on the pivot axis. The radius mismatch allows for increased space on the MMS without increasing the form factor of the entire apparatus. 

Both the hanging MMS and CMS assemblies are connected to the wheel via annealed, titanium fibers. The MMS suspension is mainly comprised of the air bearing shaft, mass pan gimbal for holding the test mass, and two coils with 3253 turns each and mean diameter of 73\,mm wound from magnet wire with a diameter of 0.06\,mm. The CMS suspension consists of a small coil and copper tube, a dual purpose deadweight and eddy current damper for suppressing the pendulum modes of the hanging assembly. The CMS coil hangs inside a closed-circuit NdFeB/mild steel magnet system.

\subsection{Air Bearing}

The MMS coils are rigidly connected to a square cross section shaft inside an air bearing housing. This linear guidance mechanism allows for almost zero-friction, single degree of freedom (DOF) translation. This design builds off of an idea introduced in \cite{Kibble} where Drs. Bryan Kibble and Ian Robinson suggest that a single degree of freedom constrained coil motion is insensitive to some misalignments, one of them being the alignment of the coil motion trajectory to the gravity vector. Simply put, if the linear trajectory of the coil is misaligned to gravity, this misalignment is present in both velocity mode and force mode, thus cancelling out. Other types of fluid bearing mechanisms have been implemented in \cite{KRISS} and \cite{MSL}.

The square air bearing is mounted on a tip/tilt stage for loosely aligning the trajectory to vertical to reduce the risk of contact friction between the air bearing shaft and the bearing surface, especially at low air pressures, and to adjust the angle of the coil inside the magnet. 

Parametric studies were conducted to examine the effect of the air bearing on the weighing mode measurement noise and bias force. The effect of the temperature of the compressed air on the weighing mode measurement bias force was studied by heating the air with a power resistor on a small copper section of the intake tube where no correlation between mass determinations and intake air temperature was found. The effect of input air pressure amplitude on measurement noise was examined and it was chosen to operate the air bearing at 41\,kPa above atmosphere, the lowest possible pressure at which the air bearing could operate before seizing because measurement noise was found to be proportional to air pressure.

\begin{figure}
\center
\includegraphics[width=3.4in]{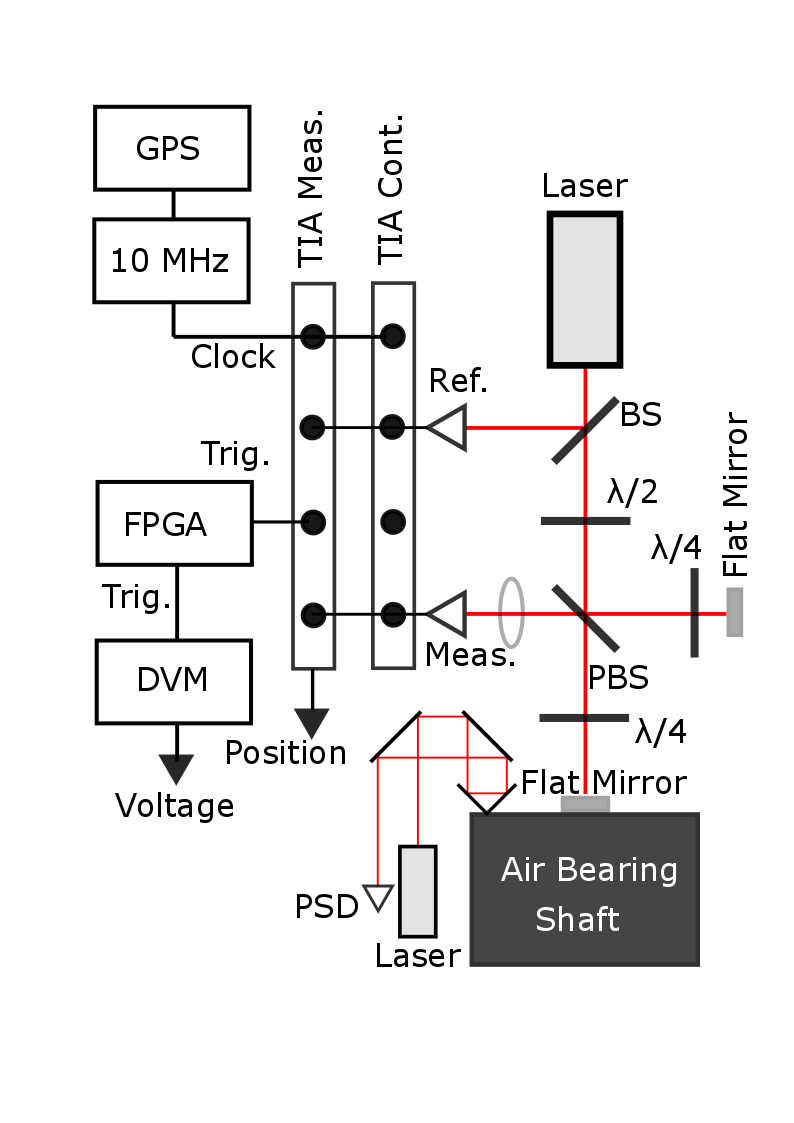}
\caption{A Michelson heterodyne interferometer measures the motion of the air bearing shaft rigidly mounted to the coil. The reference laser beam and the measurement laser beam are directed into receivers connected to two Time Interval Analyzers (TIAs), one of which continuously measures position and time and the other which measures only when triggered. A field programmable gate array (FPGA) provides the trigger pulses for both the measurement TIA and the DVM. Both TIAs are tied to a 10\,MHz reference clock synchronized with a GPS timer. A separate laser measures the horizontal motions of the air bearing.}
\label{fig:IFO}
\end{figure}

\begin{figure}
\center
\includegraphics[width=3.4in]{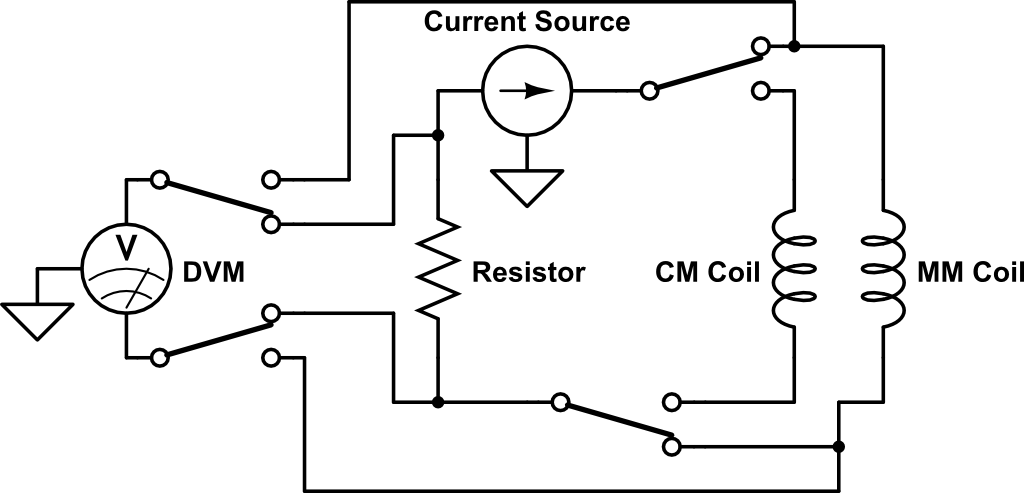}
\caption{The setup depicted is the circuit for force mode. The DVM measures the voltage drop across the resistor while the current source supplies the main mass coil. Velocity mode can be imagined if all four relays were to toggle. The DVM measures the induced voltage from the main coil while the current source drives the counter mass coil.}
\label{fig:Circuit}
\end{figure}

\subsection{Optical and Electrical}

A dual frequency 2.83\,MHz Zeeman-split laser is used as the source for the Michelson heterodyne interferometer for measuring and controlling the displacement of the main coil along $Z$. The measurement laser beam of the interferometer projects onto a flat mirror mounted centered on the top surface of the coil former adjustable in angle about $X$ and $Y$. Because the angular degrees of freedom of the coil are constrained by the air bearing, a simple flat mirror was chosen instead of a retroreflector. The reference arm projects onto a  mirror mounted to the top edge of the magnet in a similar fashion. This location was chosen to minimize the optical path difference between the two arms and for common mode rejection of mechanical vibration between the coil and magnet. The interferometer signals are read through two Carmel Instruments
\footnote{Certain commercial equipment, instruments, and materials are identified in this paper in order to specify the experimental procedure adequately. Such identification is not intended to imply recommendation or endorsement by the National Institute of Standards and Technology, nor is it intended to imply that the materials or equipment identified are necessarily the best available for the purpose.} 
time interval analyzers (TIA). One TIA serves as a continuous position and time readout for feedback control while the second TIA serves as the measurement readout for velocity only when triggered. A horizontal displacement sensor  is comprised of a separate laser beam which reflects off a corner cube mounted to the top surface of the coil former and back onto a 2D position sensor for monitoring minute parasitic $X$ and $Y$ motions of the coil during measurements.

The position and voltage measurement schematic is shown in fig.~\ref{fig:IFO}.  The two TIAs derive their timebase from a 10\,MHz clock synchronized to a GPS signal. A Field Programmable Gate Array (FPGA) deploys trigger pulses for sampling both the position and time measurements from the TIA and the voltage measurements from an Agilent 3458A Digital Voltmeter (DVM). 

The electrical circuit shown in Fig.~\ref{fig:Circuit} uses four low-noise latching relays. Critical connections are made with PTFE insulated twisted pair wires. The thin wire segments to both the MMS coils and CMS coil are made from 23\,$\upmu$m diameter wires.

\subsection{Magnetic}

\begin{figure}
\center
\includegraphics[width=3.0in]{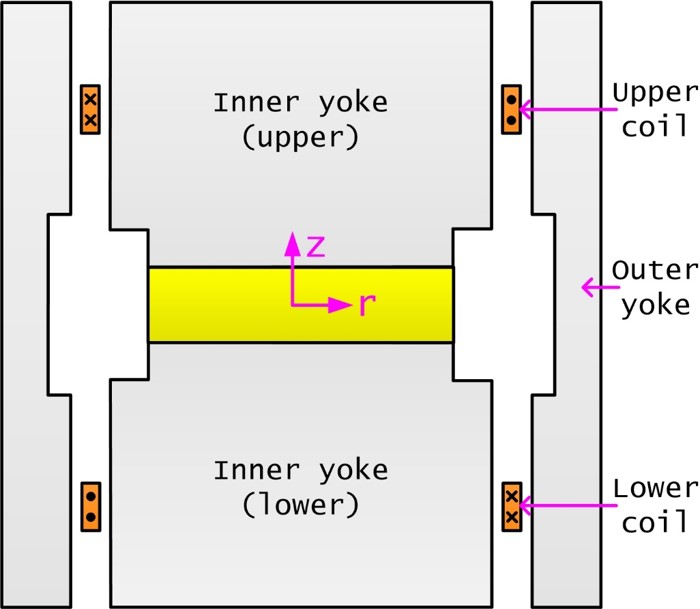}
\caption{Cross section representation of the magnet system. Two coils wound on the same former are connected in series opposition. Each coil has 3253 windings. Two halves of the outer yoke have been simplified to a single sleeve in this figure. }
\label{fig:magnetCAD}
\end{figure}

The MMS magnet system employs a single SmCo disk measuring 12.7\,mm in height and 50.8\,mm in diameter as the source of the magnetic circuit. Two nearly identical mild steel cylinders sandwich the magnet, shown in fig.~\ref{fig:magnetCAD}. These three components make up the inner yoke assembly. Two symmetric tubes made from the same steel are stacked and locked to each other via three dowel pins and serve as the outer yoke assembly. Both the inner and outer yoke assemblies are bolted to an aluminum base plate capable of tip, tilt, and vertical translation. 

\begin{figure}
\center
\includegraphics[width=3.0in]{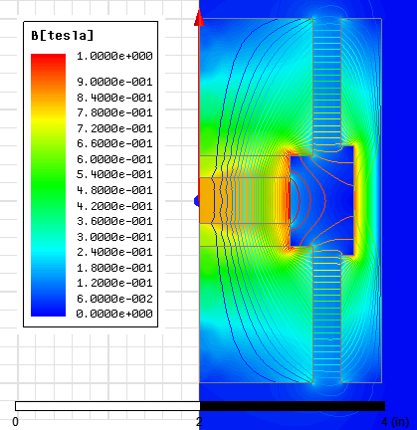}
\caption{A finite element simulation of the magnetic flux density through the top and bottom air gaps of half the magnet. The field where the coil resides in weighing mode is approximately 0.1 T.}
\label{fig:magnetFEA}
\end{figure}

\begin{figure}
\center
\includegraphics[width=3.0in]{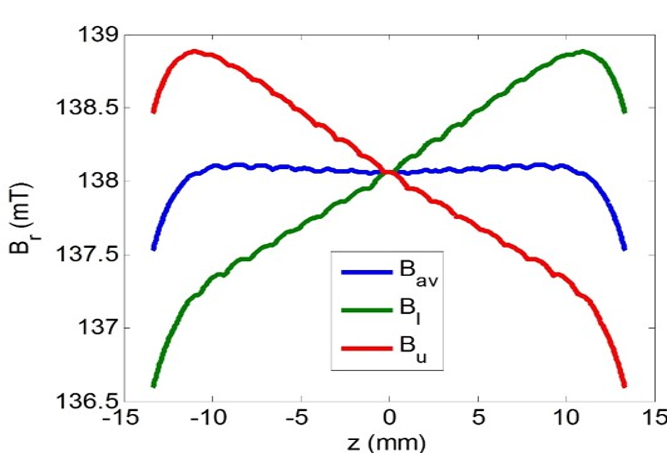}
\caption{Theoretical magnetic flux density profile of the upper air gap ($B_u$), the lower air gap ($B_l$), and the average of the two ($B_{av}$) versus vertical position of the coil $Z$. The MMS coil is comprised of two individual coils connected in series opposition so the measured profile should reflect the shape of $B_{av}$ with a local minimum at $Z=0$.}
\label{fig:BLcurves}
\end{figure}

The upper and lower 7.6\,mm wide and 35.6\,mm tall air gaps contain the radial magnetic field and are designed to guide linearly increasing or decreasing magnetic flux densities with respect to $Z$ as seen in Figs.~\ref{fig:magnetFEA}-\ref{fig:BLcurves}. Thus, in principle, the combined magnetic flux density curve is uniform in the neighborhood of $Z=0$. However, due to asymmetries of the magnet yoke pieces, a magnetic field profile with a slope of $0.7\,\times10^{-6}$/$\upmu$m near $Z=0$ was measured, as seen in fig. \ref{fig:BLsweeps}. Weighing mode controls were optimized to hold weighing position to within 0.06\,$\upmu$m and the effect of the slope was deemed to have only a minor influence on the overall uncertainty. It is possible to shape the field such that a flat spot is achieved around the weighing position. This technique is described in \cite{Chao}.

A drawback of the open top/bottom magnet system design is the leakage of the magnetic flux near the unguided regions. Therefore, any test mass will be susceptible to a systematic force from the stray magnetic field and its gradient along $Z$. The mass pan hangs approximately 50\,mm below the bottom surface of the magnet. Thus, for example, an OIML class E$_2$ 10\,g stainless steel mass placed on the KIBB-g1 mass pan would experience an additional force equivalent to a 12\,mg mass due to the magnetic susceptibility of the material. This effect can be characterized by two magnetic force terms from \cite{e1e2}:

\begin{equation}
    e_1 = \chi B_{0z}\frac{b_zV}{\mu _0}
\end{equation}

\begin{equation}
    e_2 = \mu_0 M_z\frac{b_zV}{\mu _0}
\end{equation}
where $\chi$ is the volume magnetic susceptibility of the mass, $B_{0z}$ is the absolute magnetic flux density, $\mu_0$ is the free space permeability, $M_z$ is the permanent axial magnetization, $b_z$ is the field gradient along $Z$, and $V$ is the volume of the mass. The overall force is the sum, $e_1 + e_2$. 

Both the absolute magnetic field and the field gradient were measured at the mass pan to be 4.1\,mT and 100\,$\upmu$T/mm, respectively. A field cancellation procedure of adding a 5-DOF adjustable permanent magnet underneath the mass pan to negate the field at the mass location has proven successful and in principle can reduce the magnetic field gradient to zero, thereby reducing both $e_1$ and $e_2$ to zero. However, a strong magnet placed near the mass pan is cumbersome for development purposes so we have chosen to complete our measurements with  masses made from copper where the magnetic forces are negligible. See \cite{Chao} for more information regarding figs. \ref{fig:magnetCAD} - \ref{fig:BLcurves}.

\section{Measurement Procedure and Data Analysis}

A blind measurement campaign was conducted with two weights having nominal masses of 5 g and 1 g.  The NIST Mass and Force group calibrated these weights by comparison with standards that are traceable to the SI definition of mass .  The “true mass” was determined for each weight, meaning that the raw mass readings of each weight, or apparent mass, was corrected for the effects of air buoyancy.  Furthermore, each of the weights was calibrated using a “4 -1 weighing design,” in which the unknown weight is compared with three other weights having the same nominal mass: one that serves as the “restraint” or standard mass, one that serves as a “check standard” and a third weight whose mass is unknown.  Each weight in the design is compared with every other weight, so that there are six total comparisons (mass differences) in a 4 -1 design shown in Table \ref{tab:MFgroup}.

\begin{table}[h!]
  \begin{center}
    \caption{Comparisons from mass comparator determination}
    \label{tab:MFgroup}
    \begin{tabular}{ c|c |c |c}
      \toprule % <-- Toprule here
       \textbf{W1} & \textbf{W2} & \textbf{W3} & \textbf{W4}\\
      \midrule % <-- Midrule here
      1 & -1 & 0 & 0 \\
      \midrule % <-- Midrule here
      1 & 0 & -1 & 0 \\
      \midrule % <-- Midrule here
      1 & 0 & 0 & -1 \\
      \midrule % <-- Midrule here
      0 & 1 & -1 & 0 \\
      \midrule % <-- Midrule here
      0 & 1 & 0 & -1 \\
      \midrule % <-- Midrule here
      0 & 0 & 1 & -1 \\

      \bottomrule % <-- Bottomrule here
    \end{tabular}
  \end{center}

  W1 = Standard Weigh (restraint), W2 = Check Standard Weight, W3 = Unknown Weight \#1, W4 = Unknown Weight \#2.

\end{table}

The six mass comparisons indicated in the table above form a set of six equations with two unknowns that can be easily solved using techniques from linear algebra.  Uncertainties in the restraint and check standard can be incorporated into the design along with the uncertainties in the mass comparator and air buoyancy corrections.  The result of the calculation is a mass value for each of the unknowns along with Type A (statistical) and Type B (non-statistical) uncertainties.  By incorporating a check standard, statistical tests on the measurement process and results can be performed that provide a measure of confidence. Information on the use of weighing designs in mass metrology are found in \cite{Jabbour}.

An acrylic dome is first placed over the KB to shield the instrument from air currents caused by the air conditioning in the laboratory. A small hole in the dome allows passage for the interferometer laser beam. A Vaisala PTU300 environmental sensor next to KIBB-g1 measures temperature, pressure, and relative humidity for calculating corrections for both index of refraction and buoyancy variations.

\begin{figure}
\center
\includegraphics[width=3.5in]{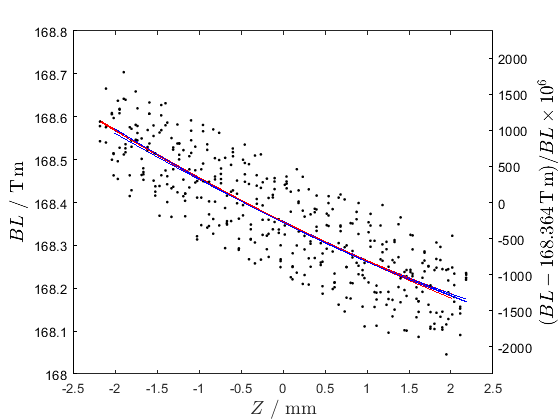}
\caption{A single set of velocity mode $BL$ measurements. Each solid line represent a second order fit of a 60-point sweep ranging between $Z = \pm\,2.2\,\mathrm{mm}$ at 1\,mm/s. In total, 6 sweeps are conducted per velocity mode measurement, where blue represents moving up and red down. Each point in fig. \ref{fig:AllBLs} represents the average value of one set of fits taken at $Z=0$. The slope around $Z=0$ is 0.118\,T\,m/mm.}
\label{fig:BLsweeps}
\end{figure}

\begin{figure}
\center
\includegraphics[width=3.5in]{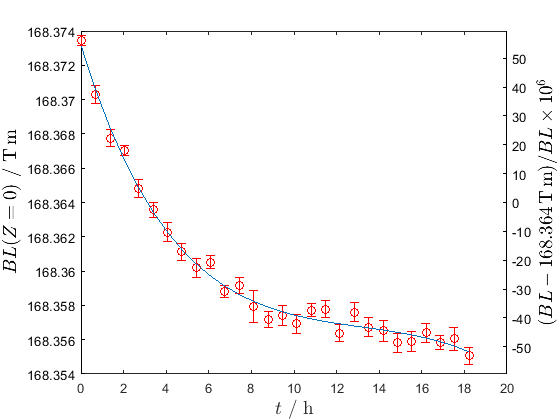}
\caption{$BL$ determinations of a full measurement spanning 18 hours for the 5\,g determination. The relative statistical uncertainty of each determination is on average $2\times10^{-6}$ ($k=1$). The drift is caused by change of the magnetization due to temperature changes of the laboratory.}
\label{fig:AllBLs}
\end{figure}

Velocity mode operates with the measurement mass resting on the mass pan and the MMS heavier by an amount equal to half the measurement mass. The measurement begins with 3 up and 3 down velocity sweeps at a constant velocity of 1\,mm/s while sampling the DVM every 1 power line cycle (NPLC) or 33\,ms with an Auto Zero in between each sampling aperture. These parameter values were chosen based on examining the power spectrum of velocity noise and a separate parametric study between differing NPLC and velocity values. The FPGA triggers both the sampling of the TIA and DVM. For example, each voltage measurement is bracketed by 17 position and time readings where each set is averaged down to a single position and time. The velocity during the voltage measurement is determined by the difference of two consecutive position readings divided by the sample time.  From the voltage and velocity data pairs the quotient is calculated; this is the $BL$. Each sweep consists of 60 $BL$ measurements ranging between $Z = \pm\,2.2\,\mathrm{mm}$ and is fitted with a second order polynomial, as seen in fig. \ref{fig:BLsweeps}. A mean and standard deviation is calculated for each  set of velocity sweeps at $Z$ = 0, represented by the red markers shown in fig. \ref{fig:AllBLs}. The entire measurement run of $BL$ determinations is fitted with a fifth order polynomial for which the interpolated values are later used in force mode calculations described below.

In between sets of velocity mode measurements, the system toggles to force mode and the coil position is servo controlled to $Z=0$, i.e., the weighing position. A motorized translation stage first removes the mass while the balance is controlled. The excursion of the balance beam from the nominal position is suppressed with large control gains. As reference, the balance excursion is approximately $\pm$40\,$\upmu$m for a 5\,g exchange. The balance then undergoes a knife-edge hysteresis erasing procedure where the balance follows a decaying sinusoidal trajectory with an initial amplitude larger than that of the perturbation caused by the mass removal. This is necessary because the knife edge is not an ideal, frictionless surface and incurs a bias restoring force depending on the direction and amplitude of the excursion from mass exchanges. After 30\,seconds settling time, 200 voltage readings caused by the weighing current across the measurement resistor are taken, once every 100\,ms.  The process is then repeated for a mass on measurement. In total, 5 mass-on and 5 mass-off measurements are taken per force mode set.

\begin{figure}
\hspace*{-0.4cm}
\includegraphics[width=3.5in]{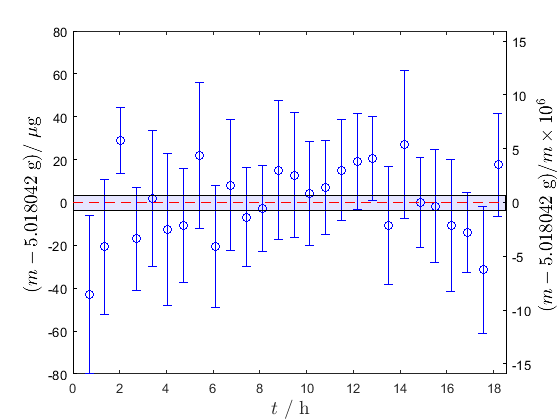}
\caption{A complete run with 27 mass determinations for a copper cylinder with a mass value  of  5.018\,042(5)\,g as determined by the Mass and Force Group. Averaging the data gives a value of 5.018\,042\,g (dashed horizontal line) with a relative statistical uncertainty of $0.7\times10^{-6}$ ($k=1$) (shaded band). The measured value was determined to be $0.03\,\times10^{-6}$ lower than the true value and this difference was rounded down to $0$.}
\label{fig:5g_finalweighs}
\end{figure}

\begin{figure}
\hspace*{-0.4cm}
\includegraphics[width=3.5in]{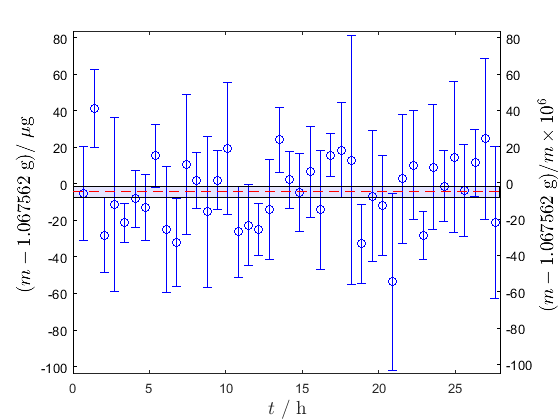}
\caption{A complete run with  41 mass determinations for a copper cylinder with a mass value of 1.067\,562(2)\,g as determined by the Mass and Force Group. Averaging the data gives a value of 1.067\,557\,g (dashed horizontal line) with a relative statistical uncertainty of $2.8\times10^{-6}$ ($k=1$) (shaded band). The measured value was determined to be $4.1\,\times10^{-6}$ lower than the true value.}
\label{fig:1g_finalweighs}
\end{figure}

Finally, the set of force mode voltage measurements are converted to 8 mass calculations via an interpolated $BL$ value from bracketing velocity mode sets. Known as A-B-A measurements, the mass calculations are defined as:

\begin{equation}
\mu_j = \frac{1}{Rg}(\frac{BL_{\textit{j}}|V_{\textit{j}}| + BL_{\textit{j+2}}|V_{\textit{j+2}}|}{2} + BL_{\textit{j+1}}|V_{\textit{j+1}}|)
\label{eq:aba}
\end{equation}
for $j = 1 ... 8$ where odd values of $j$ are mass-off measurements and even values are mass-on measurements. The balance offset is chosen such that $V_{\text{on}} \approx -V_{\text{off}}$, in other words the mass imbalance for the mass-on and mass-off measurement is approximately half the mass to be measured. The imbalance is adjusted by adding/removing wire masses on the CMS before the measurement campaign. The $BL$ value used in each term in equation~\ref{eq:aba} is a time-specific interpolated value coinciding with each mass-on or mass-off measurement, occurring once every 180 seconds. The A-B-A technique performed here removes linear drifts of the $BL$ caused by temperature changes of the magnet. The average value for a set of mass calculations is one mass determination:

\begin{equation}
m_k = \frac{1}{8} \sum_{j=1}^{8} \mu_j
\end{equation}
where $k$ represents the mass determination set number inside the full measurement run. A full measurement run of 5\,g and 1\,g mass determinations are shown in figs. \ref{fig:5g_finalweighs} -- \ref{fig:1g_finalweighs}.

The final value for the 5\,g and 1\,g mass are calculated by the unweighted mean of the mass determinations:

\begin{equation}
m_{final} = \frac{1}{n} \sum_{k=1}^{n} m_k
\end{equation}
where $n = 27$ for the 5\,g value and $n = 41$ for the 1\,g value. Their corresponding statistical uncertainties were calculated as the standard deviation of the mean:

\begin{equation}
\sigma_{final} = \frac{1}{\sqrt{n}} \sqrt{\frac{\sum_{k=1}^{n} (m_k - m_{final})^2}{n-1}}
\end{equation}

\section{Uncertainty Analysis}

The final uncertainty budgets can be found in Table\,\ref{tab:UncTable_5g}. This section serves as a reference where each error source is defined and explained, organized by category and listed in order of appearance in the uncertainty budgets. 

\subsection{Interferometer Readout}
The \textit{Laser Stability and Accuracy} term is defined as the uncertainty associated with knowing the vacuum wavelength of the laser source. This error term is based on the difference between the original wavelength measurement of the laser conducted by NIST in 2012 and a new measurement conducted in 2018. The relative difference was determined to be $5\,\times10^{-9}$.

The \textit{Deadpath Error} term is defined as the uncertainty associated with calculating the change in the differential optical path length between the reference and measurement arms of the interferometer. This differential change is caused by refractive index fluctuations and has a negligible effect. 

The \textit{Optics Thermal Drift} term is defined as the uncertainty associated with knowing the change in geometry of the optical components due to to finite coefficients of thermal expansion caused by thermal fluctuations. This effect is negligible. 

The \textit{Electronics Error} term is defined as the uncertainty associated with frequency variations at a fixed temperature of the optical receiver for plane mirror optics. This effect was calculated to be $0.1\,\times10^{-6}$ based on the performance specifications provided by the manufacturer \cite{Receiver}.

\subsection{Alignment}

The \textit{Abbe Error} term is defined as the the uncertainty associated with knowing the angular deviation caused by a horizontal separation between the measurement reflector and the trajectory axis of the air bearing. Since the mirror mount is centered on the air bearing within machining tolerances and the laser spot is centered on the mirror surface, this effect is negligible.

The \textit{Off Axis Motions} term is defined as the uncertainty associated with parasitic movements due to unwanted compliance in all DOFs except $Z$.  Horizontal displacement variations in $X$ and $Y$  were measured to be about 0.09\,$\upmu$m and 0.17\,$\upmu$m over the complete travel range, respectively. Rotation about $Y$ and $Z$ were measured during mass on and mass off weighings to be about 10\,$\,\upmu$rad and 5\,$\,\upmu$rad, respectively. Rotation about $Z$ indicated a 20\,$\,\upmu$rad jump between force mode and velocity. As reference, a cosine error between force mode and velocity mode of 1700\,$\upmu$rad will yield a $1.4\,\times10^{-6}$ error. Based on this information, the square air bearing does an excellent job of maintaining nearly zero-friction, single degree of freedom motion and off-axis errors were deemed negligible.

The \textit{Cosine Error} term is defined as the uncertainty associated with the deviation of the interferometer beam from vertical. An alcohol pool is used as a reference surface for this alignment. Angular deviations are amplified by an optical lever of 4.3\,m and the interferometer laser spot can be aligned to the alcohol pool spot to within 3\,mm. This results in a cosine error of $0.1\,\times10^{-6}$.

\subsection{Velocity}

The \textit{Timing Jitter} term is defined as the uncertainty associated with fluctuations in timekeeping by the TIA and its contribution was determined to be negligible based on the worst case scenario found in the performance specifications provided by the manufacturer \cite{TIA}.

The \textit{Wavelength Compensation} term is defined as the uncertainty associated with calculating the index of refraction correction of air. The correction has a value of approximately $264\,\times10^{-6}$ and was calculated by the modified Edlen \cite{Edlen}--\cite{Edlen2} equation:

\begin{equation}
\begin{split}
n& =  n_{tp}-[(292.75 / (t/\mbox{\SI{}{\celsius}}+273.15)] (f / \mbox{Pa})\\
           &[3.7345 - 0.0401 \times (\sigma / \upmu \mbox{m} ^{-1})]^2 \times 10^{-10}
\end{split}
\end{equation}
\\
where $n_{tp}$ is the standard air refractivity, $t$ is the temperature expressed in \SI{}{\celsius}, $f$ is the partial pressure of water vapor expressed in Pa, and $\sigma$ is the vacuum wave number expressed in $\upmu$m$^{-1}$. $f$ is a function of relative humidity and  the saturation vapor pressure calculated from air temperature, $p_{sv}$.

The lab environment is held at standard room temperature, around \SI{24}{\celsius}, 100\,kPa, and 37\,\%RH. The fluctuations on the three values were measured to be about \SI{0.03}{\celsius}, 2\,Pa, and 0.05\,\%\,RH during one set of velocity sweeps over a span of a few minutes. This yields a change in refractive index of approximately $0.02\,\times10^{-6}$. However, the uncertainty in determining the absolute index of refraction is dominated by the combined uncertainties associated with the environmental sensor's accuracy in measuring temperature, pressure, and humidity. $0.5\,\times10^{-6}$ relative uncertainty on the refractive index was calculated as a worst case scenario based on the performance specifications \cite{Vaisala}. Conducting a Monte Carlo analysis on the Edlen equation with the three inputs, each with 10,000 measured data points, reduces the uncertainty to $0.2\,\times10^{-6}$.

\subsection{Coil $Z$ Position}

The \textit{Field Gradient} term is defined as the uncertainty associated with the change in magnetic flux density as a function of $Z$ position. The measured $BL$ profile of the MMS magnet system was determined to have a $0.7\,\times10^{-6}$/$\upmu$m slope around $Z=0$ and since the position feedback controls are able to hold the weighing position to within 0.06\,$\upmu$m, the $BL$ at $Z=0$ has an uncertainty of $0.04\,\times10^{-6}$. 

The \textit{Material Thermal Expansion} term is defined as the uncertainty associated with length fluctuations of the suspended components below the measurement reflector. The weighing position is always servoed to $Z=0$ with respect to the reflector so temperature fluctuations combined with the finite thermal expansion coefficient of the coil former lead to a change in distance between the location of the coil and the measurement position. Hence, such a length change would be undetectable with the interferometer. Given the temperature variations in the room, the coil's position can be known to 0.6\,$\upmu$m which, combined with the magnetic field gradient, leads to a relative uncertainty of $0.4\,\times10^{-6}$. 

\subsection{Statistical}

Here, the \textit{Statistical} term is defined as uncertainty associated with the data analysis procedure. The final mass determination is the average value of the whole set of force mode mass determinations and has an uncertainty equal to the standard deviation of the mean. An Allan variance analysis was conducted to verify that the data was Gaussian. Each individual force mode mass determination is the average value of a group of mass calculations bracketed by two velocity mode $BL$ determinations. The uncertainty of each mass determination is the standard deviation of the mean of each group of mass calculations then added in quadrature to the uncertainty of the bracketing velocity mode measurements. These uncertainties are $0.7\,\times10^{-6}$ and $2.8\,\times10^{-6}$ for the 5\,g and 1\,g determinations, respectively.

\subsection{Profile Fitting}

The \textit{BL Interpolation} term is defined as the uncertainty associated with the calculated $BL$ value for each force mode measurement. Both a linear interpolation per every neighboring pair of $BL$ determinations and a n$^{\text{th}}$ order fit interpolation for the entire measurement were implemented and compared. The fit order is chosen depending on the drift shape of the entire measurement. A fifth order fit was selected for the 5\,g data set as shown in fig. \ref{fig:AllBLs}. The difference in the overall mass determination is $0.2\,\times10^{-6}$ between the two methods.

% How we came up with 0.2ppm: took the mean of every odd BL (BL1+BL3)/2 and looked at difference with BL2. took the mean of all the differences.

The \textit{Individual BL Profile} term is defined as the uncertainty associated with the variation of the $BL$ profile shape caused by using different fit orders per velocity mode measurement. The individual $BL$ profiles were fitted with both second and third order polynomials. The variation of the $BL$ at $Z=0$ was $0.7\,\times10^{-6}$. 

\subsection{Electrical}

The \textit{Resistor} term is defined as the uncertainty associated with knowing the absolute resistance value of a 1\,k$\Omega$ Fluke 742A resistor. This resistor was calibrated against the NIST Graphene Quantum Hall Resistor (QHR) approximately one month prior to the blind measurements\cite{Rigosi}. The resistor has been calibrated frequently over the past few years and its history indicates a relative drift of approximately $0.5\,\times10^{-6}$/year. Taking into account drift rate and the recent calibration, a conservative uncertainty of $0.1\,\times10^{-6}$ was assigned for the resistor value. 

The \textit{DVM} term is defined as the uncertainty associated with voltage measurements by a commercially available multimeter. Following best metrological practices, the Agilent 3458A used for this measurement campaign was characterized by a Programmable Josephson Voltage System (PJVS). The performance of the 1\,V range was measured, indicating a $0.4\,\times10^{-6}$ gain error and $0.02\,\times10^{-6}$ offset. %Simulating KIBB-g1 5\,g, 1\,g, and velocity mode voltages with the PJVS and measuring with the DVM indicate uncertainties of 0.3$\,\times10^{-6}$, $2\,\times10^{-6}$, and 0.3$\,\times10^{-6}$, respectively. These uncertainties were measured based on the number of sample points and sample rate equal to that used in weighing mode.

\subsection{Forces on Mass}

The \textit{Magnetic Susceptibility of Mass} term is defined as the uncertainty associated with a magnetic force between the test mass and the ambient magnetic field. This was calculated to have a relative effect of $0.02\,\times10^{-6}$.

The \textit{Balance Sensitivity} term is defined as the uncertainty associated with a torque caused by the separation between the center of mass of the wheel and and its rotational axis. The torque varies as a function of wheel angle, therefore $Z$ position of the coil. By adjusting the mass distribution of the wheel, this separation can in principle be adjusted such that the torque is constant near the weighing position. This alignment was not performed here and the balance has a 50\,$\upmu$g/$\upmu$m slope near the weighing position. In principle, the balance sensitivity comes into effect if the wheel position differs between a mass-on measurement and a mass-off measurement. This measured difference is on average 0.4\,nm, mainly due to the imperfect knife edge hysteresis erasing procedure, so a relative uncertainty of $4\,\times10^{-9}$ and $2\,\times10^{-8}$ was assigned to the 5\,g and 1\,g measurements, respectively, i.e. the effect is negligible. 

The \textit{Buoyancy} term is defined as the uncertainty associated with the calculated air buoyancy correction (i.e. density of air) for a copper mass of density 8.9\,g/cm$^3$. The correction is approximately $131\,\times10^{-6}$ with respect to the lab environmental parameters. The lab environment is held at standard room temperature, around \SI{24}{\celsius}, 100\,kPa, and 37\,\%RH. Similar to wavelength compensation, the uncertainty in determining the absolute density of air is dominated by the combined uncertainties associated with the environmental sensor's accuracy in measuring temperature, pressure, and humidity. $0.3\,\times10^{-6}$ relative uncertainty on the air desnity was calculated as a worst case scenario based on the performance specifications \cite{Vaisala}. Conducting a Monte Carlo analysis on the air density equation with the three inputs, each with 10,000 measured data points, reduces the uncertainty to $0.1\,\times10^{-6}$. 

The \textit{Balance Mechanics} term is defined as the uncertainty associated with the mechanical hysteresis of the knife edge. This effect was measured by monitoring the repeatability of weighing mode current before and after intentional knife edge excursions caused by a 5\,g mass exchange. The remaining knife edge hysteresis effects after an erasing procedure is about 1\,$\upmu$g. Thus, the relative uncertainties are $0.2\,\times10^{-6}$ and $1\,\times10^{-6}$ for the 5\,g and 1\,g measurements, respectively. 

The \textit{Gravity} term is defined as the uncertainty associated with knowing the local acceleration of gravity at the location of the test mass. The value was measured with an FG-5 absolute gravimeter next to KIBB-g1 with an uncertainty of $0.1 \times10^{-6}$, in agreement with the online Surface Gravity Prediction software value calculated by the National Oceanic and Atmospheric Administration (NOAA) gravity survey found in \cite{USGS}. However, since the FG-5 measurement was a single point determination and does not account for earth's tidal changes, the uncertainty of $g$ was inflated to $0.3 \times10^{-6}$.

The \textit{Magnet Nonlinearity} term is defined as the uncertainty associated with the level of weighing asymmetry before the start of the measurement. A weighing current-dependent offset force observed in the MMS electromagnet has a $38\,\times10^{-6}$/g effect. For example, if  $BL/g ( |I_{\mbox{off}}|- |I_{\mbox{on}}|) = $ 1\,g  then the determination would be $38\,\times10^{-6}$ lower than the true value. Prior to the start of the measurement, an iterative balancing procedure reduces the asymmetry to within 10\,mg, or a relative uncertainty of $0.4\,\times10^{-6}$.

The \textit{Air Bearing Pressure} term is defined as the uncertainty associated with input pressure dependent air bearing exhaust forces. A digital pressure gauge was connected to the intake with a resolution of 69\,Pa. Pressure data taken over 27 hours was analyzed and determined that the fluctuations coherent with the mass exchanges are limited to have a $1.1\,\times10^{-6}$ uncertainty on the 5\,g mass and $5.4\,\times10^{-6}$ on the 1\,g mass.

\begin{samepage}
\section{Uncertainty Budget}
The sources of uncertainty and their relative magnitudes for a nominally 5\,g and 1\,g mass are listed below. Entries labeled as 0.0 denote uncertainties smaller than $0.05\times10^{-6}$.

\begin{table}[h!]
  \begin{center}
    \caption{KIBB-g1 Uncertainty Budget. All uncertainties are $\times10^{-6}$}
    \label{tab:UncTable_5g}
    \begin{tabular}{l|S|l||S|l}
      \toprule % <-- Toprule here
     \textbf{Source} & \multicolumn{2}{c}{5\,g measurement} &
      \multicolumn{2}{c}{1\,g measurement}\\
      \textbf{} & \textbf{Item} & \textbf{Subtotal}&
        \textbf{Item} & \textbf{Subtotal} \\
      \midrule % <-- Midrule here
      
            Laser Stability/Accuracy & 0.0 &&0.0 \\
      Deadpath Error & 0.0 &&0.0 \\
      Optics Thermal Drift & 0.0 & &0.0\\
      Electronics Error & 0.1 &&0.1 \\
      \,\,\,\,\,\,\textbf{Interferometer Readout} &  & \textbf{0.1}&& \textbf{0.1} \\
      \midrule % <-- Midrule here
      
           Abbe Error & 0.0 &&0.0 \\
      Off Axis Motions & 0.0 &&0.0 \\
      Cosine Error & 0.1 &&0.1 \\
      \,\,\,\,\,\,\textbf{Alignment} &  & \textbf{0.1}&& \textbf{0.1} \\
      \midrule % <-- Midrule here
     
           Timing Jitter & 0.0 & &0.0\\
      Wavelength Compensation & 0.2 &&0.2 \\
      \,\,\,\,\,\,\textbf{Velocity} &  & \textbf{0.2}&&\textbf{0.2}\\
      \midrule % <-- Midrule here
     
           Field Gradient & 0.0 &  &0.0\\
      Material Thermal Expansion & 0.4 & &0.4\\
      \,\,\,\,\,\,\textbf{Coil Z Position} &  & \textbf{0.4} && \textbf{0.4}\\
      \midrule % <-- Midrule here
     
           \,\,\,\,\,\,\textbf{Statistical} &  & \textbf{0.7}&& \textbf{2.8} \\
      \midrule % <-- Midrule here
     
           BL Interpolation & 0.2 &&0.2 \\
      Individual BL Profile & 0.7 &&0.7 \\
      \,\,\,\,\,\,\textbf{Profile Fitting} &  & \textbf{0.7}&& \textbf{0.7} \\
      \midrule % <-- Midrule here
     
           Resistor & 0.1 & &0.1\\
      DVM (Force Mode) & 0.4 & & 0.4\\
      DVM (Velocity Mode) & 0.4 & & 0.4\\
      \,\,\,\,\,\,\textbf{Electrical} &  & \textbf{0.8} &&\textbf{0.8}\\
      \midrule % <-- Midrule here
     
           Magnetic Susc. of Mass  &  0.0& &0.0& \\
                 Balance Sensitivity     &  0.0& &0.0& \\
      Buoyancy                &  0.1& &0.1& \\
      Balance Mechanics       &  0.2& &1.0& \\
      Gravity                 &  0.3& &0.3& \\      
      Magnet Nonlinearity     &  0.4& &0.4& \\
      Air Bearing Pressure    &  1.1& &5.4&  \\      
      \,\,\,\,\,\,\textbf{Forces on mass} &  & \textbf{1.2}&&\textbf{5.5}\\
      \midrule % <-- Midrule here

      \,\,\,\,\,\,\textbf{Total} &  & \textbf{1.8}&& \textbf{6.3} \\      
      \bottomrule % <-- Bottomrule here
    \end{tabular}
  \end{center}
\end{table}

\end{samepage}

\section{Results and Discussion}
Analyzing the KIBB-g1 5\,g and 1\,g blind measurement data yielded the following values with their associated uncertainties ($k = 1$) shown in Table \ref{tab:UncTable}. These data are compared to mass measurements conducted with the conventional subdivision method scaling from a 1\,kg artifact and shown in Table\,\ref{tab:UncTable}. $\Delta m$ is the difference in $\upmu$g between the two different measurement types and the absolute uncertainties are added in quadrature. $\Delta m/m \times 10^6$ is the same difference but expressed in relative terms, parts in $10^6$. Table\,\ref{tab:E1E2KibbTable} shows a comparison of the max permissible relative uncertainties of class E$_2$ weights with the KIBB-g1 relative uncertainties presented in this paper.

\begin{table}[h!]
  \begin{center}
    \caption{Comparison of Conventional Mass Dissemination vs. KIBB-g1 Mass Realization }
    \label{tab:UncTable}
    
    \begin{tabular}{l|p{10mm}p{1mm}p{6mm}|p{10mm}p{1mm}p{5mm}}
    \toprule % <-- Toprule here
    & \multicolumn{3}{c|}{\textbf{5\,g mass}} & \multicolumn{3}{c}{\textbf{1\,g mass}}\\
    \midrule
    
    Conventional
    & $5.018\,042\,$g&$\pm$&$0.5\,\upmu$g 
    & $1.067\,562\,$g&$\pm$&$0.1\,\upmu$g \\
    KIBB-g1
    & $5.018\,042\,$g&$\pm$&$9.0\,\upmu$g  
    & $1.067\,557\,$g&$\pm$&$6.7\,\upmu$g \\
    \midrule
    $\Delta$m
    & $0$\,$\upmu$g &$\pm$&$9.0\,\upmu$g 
    & $5$\,$\upmu$g &$\pm$&$6.7\,\upmu$g \\
    $\Delta$m/m  $\times 10^6$
    & $0$ &$\pm$&$1.8$
    & $5$&$\pm$&$6.3$\\
   \bottomrule % <-- Bottomrule here
\end{tabular}
\end{center}
\end{table}

\begin{table}[h!]
  \begin{center}
    \caption{Comparison of E$_2$ Mass vs. KIBB-g1 Mass Realization  Relative Uncertainties}
    \label{tab:E1E2KibbTable}
    \begin{tabular}{ c|c |c }
      \toprule % <-- Toprule here
       & \textbf{5\,g mass} & \textbf{1\,g mass} \\
      \midrule % <-- Midrule here
      $\Delta$m$_{E2}/$m$_{E2}\times 10^6$ & $2$ & $5$ \\
      \midrule % <-- Midrule here
      $\Delta$m/m\,$\times 10^6$ & $1.8$ & $6.3$ \\
      \bottomrule % <-- Bottomrule here
    \end{tabular}
  \end{center}
\end{table}

\begin{figure}
\hspace*{-0.4cm}
\includegraphics[width=3.9in]{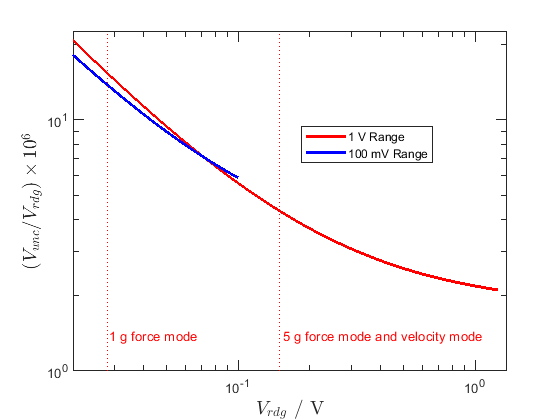}
\caption{Voltage readings and their associated relative uncertainties as specified by the Agilent 3458A manual. Both the 1\,V and 100\,mV ranges are plotted. The left dotted line indicates the voltage level measured for a 1\,g determination and the right dotted line indicates the voltage level measured for 1\,mm/s velocity sweeps and a 5\,g determination.}
\label{fig:DVMunc}
\end{figure}

Also worth noting is that various techniques are available to achieve slightly worse uncertainties in the voltage measurement without access to a PJVS simply by using manufacturer specifications which have been shown to err on the conservative side \cite{DVM}. The voltage measurement uncertainties are calculated as $4\,\times10^{-6}$, $14\,\times10^{-6}$, and $4\,\times10^{-6}$ for KIBB-g1 5\,g, 1\,g, and velocity mode measurements, respectively, shown in fig. \ref{fig:DVMunc}. 

The $14\,\times10^{-6}$ voltage uncertainty estimated from the manufacturer specifications for the 1\,g measurement is due to sampling near the bottom of the 1\,V range. Using a 10\,k$\Omega$ instead of a 1\,k$\Omega$ resistor to convert the weighing current into a voltage would reduce this estimate of the 1\,g force mode voltage uncertainty term to $3\,\times10^{-6}$. Further optimization of the magnet and coil designs to be considered for the next version can reduce all voltage measurement uncertainties to approximately $2\,\times10^{-6}$ based on the manufacturer specifications.

For laboratories with access to well-characterized Zener references, one may consider frequently calibrating their DVM for a lower uncertainty compared to the above method. For even higher accuracy measurements without a quantum standard, one may consider building a fixed resistive divider connected to the Zener and using KIBB-g1 as a fixed point mass determination device.

As a prototype tabletop instrument, KIBB-g1 has shown promise for directly realizing gram-level masses in air with uncertainties on the order of a few parts in $10^6$ and the overall measurement uncertainties are competitive with E$_2$ mass standards. Four of the five design goals mentioned at the beginning of this paper have been met; the last goal will be achieved once similar levels of uncertainties can be reached without the costly PJVS system. A rigorous study is in the works to optimize voltage measurement uncertainties while minimizing reliance on a Zener reference and/or PJVS to drive down the operating costs. In the near future, we plan to improve the mechanics with a focus on ergonomics and ease-of-use. The next iteration will bring us much closer to KIBB-g1 transitioning from a national metrology lab grade instrument to a ubiquitous weighing device.

\section*{Acknowledgment}
The authors would like to thank Shisong Li for the magnet design, Alireza Panna for the resistor calibration, Bryan Waltrip and Mike Berilla for the current source design, Stefan Cular for the voltage calibration, and Patrick Abbott, Donna Kalteyer and Kevin Chesnutwood for the mass calibrations.

\end{document}